\documentclass{article}
\usepackage[utf8]{inputenc}

\usepackage{graphicx}
\usepackage{lipsum}
\usepackage{amsmath}
\usepackage{amssymb}
\usepackage{float}

\usepackage{hyperref}

\usepackage{aaai20}

\title{ColosseumRL: A Framework for Multiagent Reinforcement Learning in $N$-Player Games}
\author{Alexander Shmakov, John Lanier, Stephen McAleer, \\
\Large \textbf{Rohan Achar, Cristina Lopes, and Pierre Baldi} \\ \\
University of California Irvine, Donald Bren School of Information and Computer Sciences \\
\{ashmakov, jblanier, smcaleer, rachar\}@uci.edu, \{lopes, pfbaldi\}@ics.uci.edu
}
\date{October 2019}

\begin{document}

\maketitle

\begin{abstract}
Much of recent success in multiagent reinforcement learning has been in two-player zero-sum games. In these games, algorithms such as fictitious self-play and minimax tree search can converge to an approximate Nash equilibrium. While playing a Nash equilibrium strategy in a two-player zero-sum game is optimal, in an $n$-player general sum game, it becomes a much less informative solution concept. Despite the lack of a satisfying solution concept, $n$-player games form the vast majority of real-world multiagent situations. In this paper we present a new framework for research in reinforcement learning in $n$-player games. We hope that by analyzing behavior learned by agents in these environments the community can better understand this important research area and move toward meaningful solution concepts and research directions. The implementation and additional information about this framework can be found at \href{https://colosseumrl.igb.uci.edu/}{https://colosseumrl.igb.uci.edu}.
\end{abstract}

\section{Introduction}
Recently, reinforcement learning has had success beating humans in 2-player zero-sum games such as Go \cite{silver2018general}, Starcraft \cite{vinyals2019grandmaster}, and Poker \cite{brown2018superhuman}. Two-player zero sum games have many nice properties that make training a reinforcement learning agent for them relatively straightforward. Mainly, playing a Nash equilibrium strategy in zero-sum two-player games is optimal, and in the worst case results in a tie. This fact forms the basis of most current algorithms in this domain, such as Monte Carlo tree search \cite{browne2012survey} and counterfactual regret minimization \cite{zinkevich2008regret}. 

While most empirical and theoretical success has come in zero-sum two-player games, we rarely find examples of these games in practical problems. Much more often we have scenarios with more than two agents, and where one agents success is not necessarily another agents failure. These $n$-player general sum games include scenarios such as self-driving cars, marketplaces, robotic teams, and the global economy. Unfortunately, the Nash equilibrium solution concept does not provide a satisfying solution to these $n$-player zero sum games. Besides being intractable to compute, a Nash equilibrium strategy has each player playing  a best response to all the other players in that current strategy. Another way to put this is that no player has incentive to unilaterally deviate. However, if two players deviate at the same time, they could be better off than if they hadn't. For example, in three-player Kuhn poker, one Nash equilibrium has the first two players getting negative utility and the third player getting positive utility. Neither the first or second player is incentivized to unilaterally deviate, but if they both deviate they could end up with positive utility. Furthermore, even if every player plays a Nash equilibrium strategy, their combined strategy might not be a Nash equilibrium strategy. Currently, there is no satisfying theoretical solution concept in $n$-player general sum games. 

Although the theory behind $n$-player general sum games is lacking, in practice many of the same algorithms that find success in two-player zero-sum games such as CFR, tree search and self-play can find very good strategies in $n$-player general sum games. Perhaps the most well-known example of this is Pluribus \cite{brown2019superhuman}, the algorithm that beat the top human players in multiplayer poker using the same techniques from the two-player superhuman algorithm Libratus. Pluribus is conjectured to be highly exploitable, but it is still able to beat the best humans at multiplayer poker. How is this possible? What can we learn from solutions found from traditional methods when used in the $n$-player general sum setting? We believe that these questions need to be emphasized in the multiagent reinforcement learning community. To that end, we have created a framework for reinforcement learning in $n$-player general sum games. This framework includes a number of new $n$-player game environments such as Blokus, Tron, and 3/4 player tic-tac-toe as well as traditional normal form games and Kuhn poker. Our hope is that by comparing learning approaches and algorithms against each other in these environments we can better understand the dynamics of $n$-player games and start to discover relevant solution concepts. 


\section{ColosseumRL}
We introduce a multi agent framework designed to allow researchers to tackle the challenges presented. Colosseum is a reinforcement learning framework that includes several environments, a unified format for adding new environments, and a distributed competition server for testing and evaluation. The matchmaking system allows for multi agent competitions to be executed in a distributed manner, allowing the evaluation of large populations of agents.

\section{Environments}
Currently, Colosseum has three implemented competitive $n$-player environments. We also define a generic framework for defining $n$-player games with potential imperfect information that will automatically integrate with the evaluation features.

\subsection{Tic-Tac-Toe} 
\begin{figure}[h]
\centering
\includegraphics[width=0.95\columnwidth]{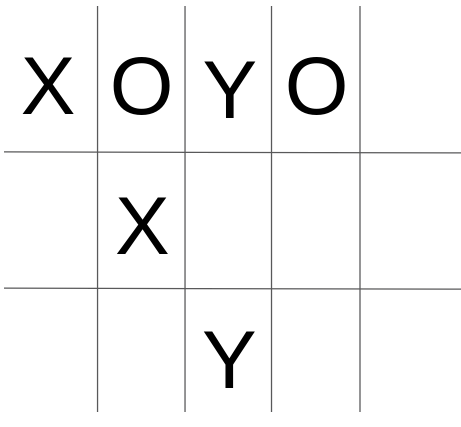}
\end{figure}
We extend the notion of Tic-Tac-Toe to $n$-player games by using identical rules to regular tic-tac-toe and introducing additional symbols for players. We create an $n \times m$ board use the symbols $X, O, Y, Z, \ldots$ in order to label the players. The goal for every player is to attain three of their symbols sequentially in any cardinal or inter-cardinal direction. The winner of the game is the player that achieves this configuration first, and all of the other players are equal losers. Rewards for this game are $r = 0$ for a regular action, $1$ for winning, and $-1$ for losing. The observation for this game is the entire game board, with no obscured information. This game is a perfect information game with a relatively small game tree, meaning that agents may search essentially all possibilities given enough computation time. Tic-Tac-Toe represents a good challenge for multi-agent reinforcement learning due to the presence of king-maker scenarios: states of the game where one player has no possibility of winning but they may decide which of the other two players would win. This challenges tree search algorithms because it undermines the notion of optimal play in all scenarios.

\subsection{Blokus} 
\begin{figure}[h]
\centering
\includegraphics[width=0.95\columnwidth]{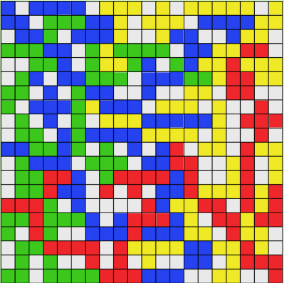}
\end{figure}
Blokus is a strategy board game where the players' goal is to gain maximum control over the board. Blokus has a maximum of 4 players, each receiving one of every possible polyominos ranging one to five sections each. Players may place these polyominos in any diagonally adjacent location to the pieces they already have. Every time a player places a block, they gain control over the number of sections that block contained; for example, large pentonimos give five control whereas the smaller dominos only give two. Rewards for this game are equal to the amount of control received for each move, or $0$ if no more moves are possible. At the end of the game, players are given a ranking from first to fourth based on their total control. The observations for this environment are the entire board and the pieces each player has remaining; and, since it is a sequential board game, it is again fully observable with perfect information. Although Blokus only has a maximum of four players, it posses interesting challenges for reinforcement learning due to the area control aspect of the game. It is easy for players to form short term alliances by focusing their control efforts on a single player and avoiding blocking each other's possible moves. Also, although the game is perfect information and sequential, the large action spaces makes simple search algorithm infeasible for finding optimal agents.

\subsection{Tron} 
\begin{figure}[h]
\centering
\includegraphics[width=0.95\columnwidth]{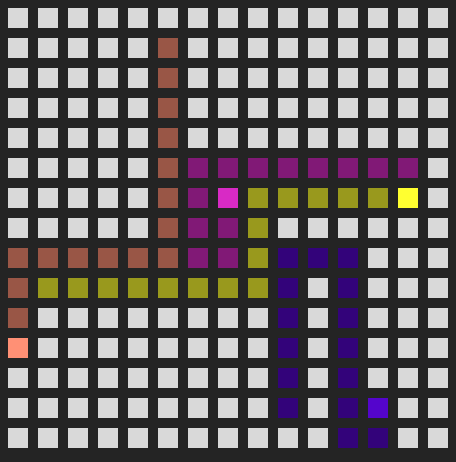}
\end{figure}
Tron is simple $n$-player grid based video game. The game consists of an $K \times M$ arena with surrounding walls and $N$ players initially starting in an circle around the arena. Players may move forward, left, or right and when doing so, leave behind a trailing wall. Players are eliminated if they crash into either the arena wall, another player's wall, or their own wall. The goal of the game is to force your opponents to crash into a wall by limiting their movement options. The game supports arbitrarily high number of players as well as teams of multiple players. The rewards are $1$ for surviving an action, $-1$ for crashing, and $10$ for achieving first place without a tie. At the end of the game, players are ranked (with possible ties) based on the order that they crashed. The observation for Tron may either be the entire game board or a small window around only the given player. Tron may also either be played sequentially for a perfect information variant of the game or it can be played simultaneously to mimic the real game. Tron is an interesting game to study MARL on because it is easily extendable to many players, features an area control mechanic that allows players to form short-term alliances, and allows for a large of amount of simple variations.

\subsection{Normal Form Games}
\subsubsection{Random Payoff Matrix}
For theoretical exploration, we implement games with randomly generated payoff matrices. We define each environment with a payoff matrix $M \in \mathbb{R}^{S_1, S_2, \ldots, S_P, P}$, where $P$ is the number of players and $S_i$ is the number of possible actions for player $i$. The payouts for each action are randomly generated from a uniform distribution on $[0, 1)$, and we include an option to ensure that the game will always be zero-sum. The observation for this environment is the entire normal form game matrix, and the action for each agent is a number from $0$ to $S_i$.

\subsubsection{Rock Paper Scissors}
A classical environment from game theory is Rock Paper Scissors, which servers as a common baselines for evolutionary mechanics. We provide both the traditional two player rock paper scissors as well as a generalization of the game to three players. This generalization, created by Bernard Koven \cite{three_player_rps}, extends the rules so that the winner of any game is the player with the single unique action out of the three players. If all three players have the same action or all three are unique, then the game is a tie. We allow for user-defined payoff values for winning, losing, and tying the game. The observation for this environment is the normal form matrix given the payoff values, and the actions for each players are the traditional Rock, Paper, and Scissors.

\subsubsection{Kuhn Poker}
A classical example of a partial information environment from game theory is Kuhn Poker, a simplified version of Texas Hold'em. The original version of the game designed by \cite{kuhn} and involves two players with a maximum of three actions in a game. A generalization of the mechanics to three players was defined and studied by \cite{three_player_kuhn}. We further generalize the mechanics to include an arbitrary $N$ players, although the analogy to Texas Hold'em breaks down with enormous player counts. The deck of $N$ player Kuhn Poker consists of $N + 1$ cards labeled $0$ to $N$. The game consists of two rounds: If there is no outstanding bet, then a player may either bet or check. If there is an outstanding bet, then a player may either call or fold. If all players have checked or a player has bet and at least one other has called, then the participating player with the highest card wins the entire pot. Reward is 0 for each action and the change in net worth at the end of a game. Three player Kuhn poker is one of the few non-trivial multiplayer games with known Nash equilibria.

\section{Baselines}
One major difficulty in multi agent reinforcement learning is finding good opponents to collect episodes from. We may only train a competent agent if we have good opponents to train on. We apply several standard multiplayer reinforcement learning techniques to examine the difficulties of training in these MARL environments. All of the agents are optimized using Proximal Policy Optimization \cite{schulman2017proximal} using the same neural network, and we only vary the training strategy. We focus on the simultaneous-action fully-observable Tron environment for these test because it has a small enough action and observation space for a network to reliably learn but that is not trivial enough for tree search. We use a tron board with size $15 \times 15$ and $4$ players.

\subsection{Fixed Opponent}
We first train an agent using single player reinforcement learning against a simple, hand-coded agent. This agent moves forward if the path is clear, or randomly moves either left or right if the path is clear in either direction. We train a neural network on this fixed opponent to ensure that our PPO and network are capable of learning the environment and task given a stationary opponent. We test three different variants of these fixed agents. In each variant, we change the probability that the agent will perform a random action instead of the hand-coded action. We set this probability to 5\% ($S\alpha$), 25\% ($S\beta$), and 100\% ($S\gamma$).

\subsection{Self Play}
We also apply the common two-player tactic of delayed update self play (SP). In this technique, we train against fixed copies of ourselves until the current agent reaches a specified average win percentage. We then update all of the opponent agents and repeat this training procedure on the new agents. This will allow the agent to learn against better opponents than our hand-coded option, but it also provides stability since we do not update the opponents very often. We find that self play converges to a stable agent within 1 to 2 million time-steps. We try two different minimum win percentages for the update: 50\% ($SP\alpha$) and 80\% ($SP\beta$). Lower update percentages produces quicker updates to the opponents.

\subsection{Fictitious Self Play}
We also use an extended version of self play known as fictitious self play \cite{heinrich2015fictitious}. In this variant, instead of only keeping track of the last iteration of the target network, we keep the last $K$ iterations and play against all of them. In each game, each opponent has an 80\% chance of being the latest iteration of the policy and a 20\% change of being one of the other $K - 1$ iterations. This encourages diversity in the opponent population and ensures that our agent does not over-fit to itself. We again try update percentages of 50\% and 80\% and population sizes of $K = 4$ ($FSP\alpha$, $FSP\beta$) and $K = 16$ ($FSP\gamma$, $FSP\delta$). Higher populations should create more robust agents since they have to maintain dominance against a wider variety of opponents.

\section{Evaluation}
Another major difficulty in multi agent reinforcement learning is the task of ranking a population of trained agents. Simple round robin tournaments may results in non-transitive relations between agents. It may be difficult to find a single agent who beats all others, but simply looking at each agents win count may not always find the intuitively best agent either. 

\begin{figure}
    \centering
    \textbf{Distribution of Rankings}\par\medskip
    \includegraphics[width=0.48\textwidth]{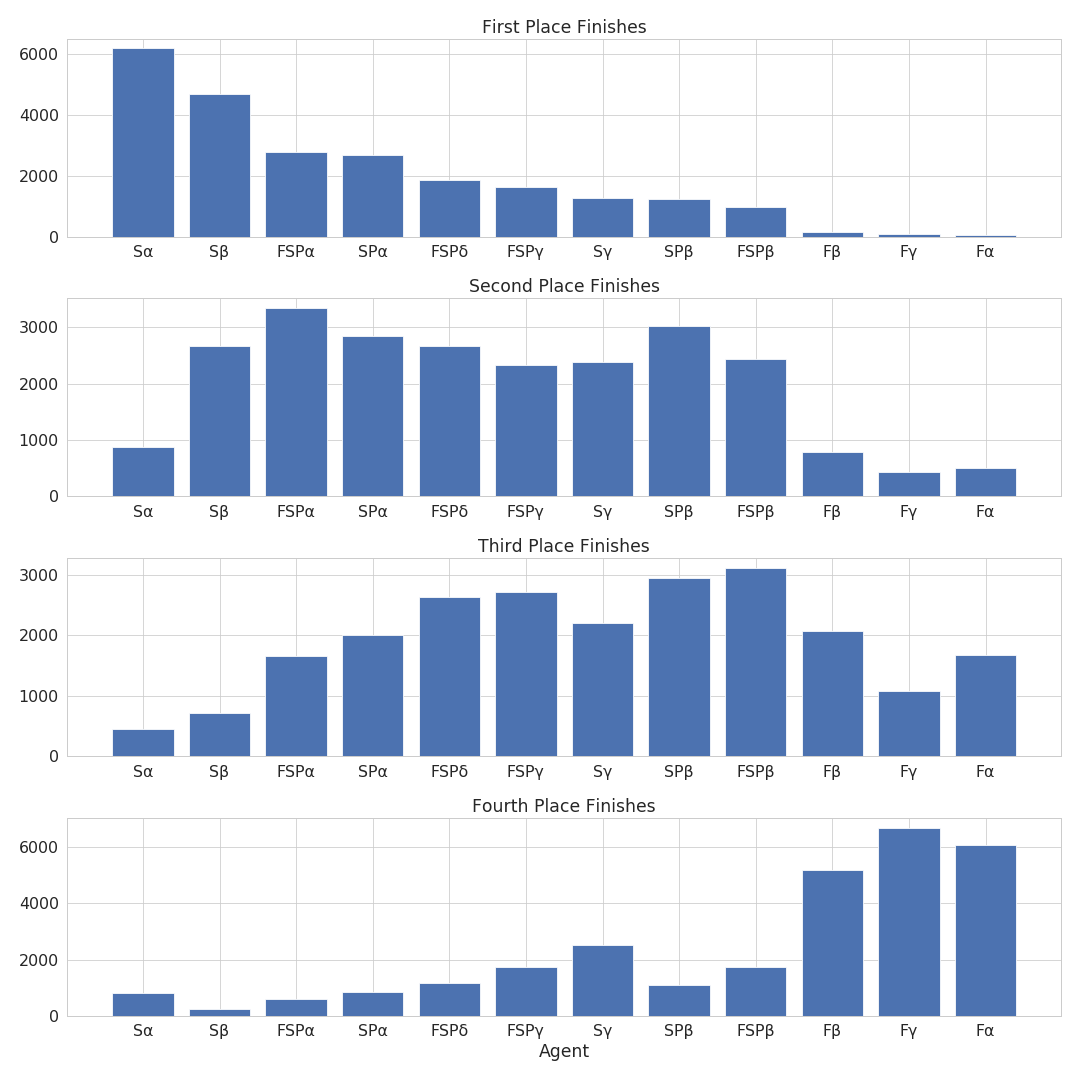}
    \caption{The distribution of rankings each agent received throughout tournament.}
    \label{fig:results}
\end{figure}

We provide a simple measure of agent performance based on the probability of an agent finishing in a given position. We run a tournament of $25,000$ games, in which we uniformly sample the opponents from our 12 variations without replacement. From every game, we receive a ranking for each agent based on when they crashed into a wall. We collect the number of times each agent finished first, second, third, and fourth. We then sort these agents based on their first place performance and plot the times each agent achieved each ranking. 

For model design, we flatten our board representation and stack the last three observations so the agent has access to temporal information about the environment. We then feed this flattened observation into a two layer feed forward network with hidden sizes of $512$ and $256$. Finally, we split this hidden vector into a policy and value layer. We train each agent for 2.72 million timesteps with $5,440$ timesteps for each training iteration. Our network runs 16 iterations of of the Adam Optimizer with an SGD batch size of 1024 for each training batch.

We run a tournament featuring the three agents trained on those fixed opponents, the two self-play agents, and the four variations of fictitious self play agents. We also include three variants of our fixed agent to act as a sanity check for the other agents ($F\alpha$, $F\beta$, $F\gamma$). We note that if two players tied, for example if they both crashed into a wall at the same time, their rank is rounded down to the lowest nearest rank.

The results of the competition are showing in Figure \ref{fig:results}. We see an interesting result that the agents trained against only the fixed opponent with single player reinforcement learning performed the best. The self play and fictitious self play policies performed less good, and the fixed opponents were always last. This is a preliminary analysis, and raises more questions than it answers. 

As an additional test, we ranked the agents using the ranked pairs voting scheme \cite{tideman1987independence}. Ranked pairs constructs an acyclic directed graph by sequentially placing directed edges between agents in order of their win percentage, checking that each edge does not create a cycle. The rankings generated by this voting scheme, in order from highest to lowest, are as follows: $\{ S \alpha, S\beta, FSP\alpha, SP\alpha, FSP\delta, SP\beta, FSP\gamma, S\gamma, FSP\beta, \\ F\beta, F\alpha, F\gamma \}$. 

\section{Discussion}
It is striking that even in the simplest possible $n$-player games such as three-player tic-tac-toe, we don't know what an optimal strategy for an individual player is. This is concerning because $n$-player games form the vast majority of real-world multiagent systems. We believe that the multiagent reinforcement learning community needs to put much more emphasis on studying these games and designing algorithms that learn how to play optimally in these games. We also think that much greater emphasis needs to be placed in this area on the theoretical side from game theorists, both in defining ranking and optimality concepts as well as developing algorithms that create agents which achieve them. 

In this paper we have mainly focused on the question of how to design an optimal agent or rank agents in general sum $n$-player games. Perhaps this is the wrong question though. Maybe in real-world $n$-player scenarios, finding the optimal strategy for each agent doesn't make sense. For example, in self-driving cars, we don't want to have each car get to their location the fastest by endangering other cars. There is a large literature in economics on social choice theory, which approaches this problem from the perspective of a social planner. It is possible that this is a more useful direction because the price of anarchy can be quite large in $n$-player games. There is also much work done on cooperative $n$-player games which might be helpful in this area. Finally, it is interesting to examine connections to other $n$-player settings that are well-studied in economics such as market design and mechanism design.  



\bibliography{marl.bib}
\bibliographystyle{aaai}

\end{document}